\documentstyle[preprint,aps,epsfig]{revtex}
\def\tell{\tilde{e}}
\def\tnu{\tilde{\nu}}
\begin{document}
\draft
\preprint{\hfill ANL-HEP-CP-98-93}
\title{SUSY Production Cross Sections}
\author{Edmond L. Berger$^a$, Brian Harris$^b$, Michael Klasen$^a$, and 
Tim Tait$^{a,c}$}
\address{$^a$High Energy Physics Division,
             Argonne National Laboratory \\
             Argonne, Illinois 60439 \\
$^b$Physics Department, Florida State University, Tallahassee, Florida, 32306\\
	$^c$Michigan State University, East Lansing, Michigan 48824}
\date{\today}
\maketitle

\begin{abstract} 
We summarize the status of next-to-leading order perturbative quantum 
chromodynamics (pQCD) calculations of the cross sections for the production of 
squarks, gluinos, neutralinos, charginos, and sleptons as a function of the 
produced sparticle masses in proton-antiproton collisions at the hadronic 
center-of-mass energy 2~TeV.
\end{abstract}
\vspace{0.2in}

\section{Perturbative QCD Results}
\label{sec:1}

The possibility of supersymmetry (SUSY) at the electroweak scale and 
the ongoing search for the Standard Model (SM) Higgs boson constitute two 
major related aspects of the motivation for the Tevatron upgrade currently 
under construction at Fermilab.  The increase in the center-of-mass energy to 
2~TeV and the luminosity to an expected 2 fb$^{-1}$, together with detector
improvements, should permit discovery or exclusion of supersymmetric partners 
of the standard model particles up to much higher masses than at 
present~\cite{torah}. 

Estimates of the production cross sections for pairs of supersymmetric 
particles may be computed analytically from fixed-order 
quantum chromodynamics (QCD) perturbation theory.  Calculations that include 
contributions through next-to-leading order (NLO) in QCD have been performed 
for the production of squarks and gluinos \cite{Hopker}, top squark pairs 
\cite{Plehn}, slepton pairs \cite{Baer}, gaugino pairs \cite{Klasen}, and the 
associated production of gauginos and gluinos \cite{BKT}. The cross 
sections can be calculated as functions of the sparticle masses and mixing 
parameters. 

In a recent paper \cite{Berger}, Berger, Klasen, and Tait provide numerical 
predictions at next-to-leading order for the production of squark-antisquark, 
squark-squark, gluino-gluino, squark-gluino, and top squark - antitop squark 
pairs in 
proton-antiproton collisions at the hadronic center-of-mass energy 2~TeV.  
These calculations are based on the analysis of Refs. \cite{Hopker,Plehn}, 
and the CTEQ4M parametrization \cite{CTEQ4} of parton densities.   The hard 
scale dependence of the cross section at leading order (LO) in perturbative 
QCD is reduced at NLO but not absent.  An estimate of the theoretical 
uncertainty at NLO is approximately $\pm 15$~\% about a central value.  
The central value is obtained 
with the hard scale chosen to be equal to the average of the masses of the 
produced sparticles, and the band of uncertainty is determined from a 
variation of the hard scale from half to twice this average mass.  The 
next-to-leading order contributions increase the production cross sections 
by 50~\% and more from their LO values.  For example, in the case of 
squark-antisquark production the next-to-leading order cross section lies 
above the leading order cross section by 59~\%.  This increase translates into 
a shift in the lower limit of the produced squark mass of 19~GeV.  The cross 
sections for squark-antisquark production, gluino pair production, and the 
associated production of squarks and gluinos of equal mass are of similar 
magnitude, whereas the squark pair production and top squark-antitop squark 
production cross sections are smaller by about an order of 
magnitude \cite{Hopker,Plehn}.

The cross sections reported in Ref.~\cite{Berger} are for inclusive yields, 
integrated over all transverse momenta and rapidities.  In the search for 
supersymmetric states, a selection on transverse momentum will normally be 
applied in order to improve the signal to background conditions.  The 
theoretical analysis can also be done with similar selections.  A tabulation 
of cross sections for various squark and gluino masses is available upon 
request from the authors of Ref. \cite {Berger}.

Next-to-leading order calculations of the production of neutralino pairs, 
chargino pairs, and neutralino-chargino pairs are reported to be 
on the way to completion \cite{Klasen}, but final numerical predictions are 
not yet available for general use.  

The strongly interacting squarks and gluinos may also be produced singly 
in association with charginos and neutralinos.  Leading-order production cross 
sections for the associated production of a chargino plus a squark or gluino 
and of a neutralino plus a squark or gluino are published~\cite{assoclo}, and 
a next-to-leading order calculation of associated production of a gaugino plus 
a gluino is now available~\cite{BKT}.  

Berger, Klasen, and Tait~\cite{BKT} compute total cross sections for all the 
gaugino-gluino production reactions $\tilde g \tilde{\chi}^0_{(1-4)}$ and 
$\tilde g \tilde{\chi}^{\rm \pm}_{(1-2)}$ in next-to-leading order SUSY-QCD.  
For numerical results, they select an illustrative mSUGRA scheme in which the 
GUT scale common scalar mass $m_0 = 100 $ GeV, the common gaugino mass 
$m_{1/2} =150 $ GeV, the trilinear coupling $A_0 = 300 $ GeV, 
$\rm{tan}(\beta) = 4$ and $\rm{sgn}(\mu) = +$.  (The sign convention for 
$A_0$ is opposite to that in the ISASUGRA code).  They convolute the NLO hard 
partonic cross sections with the CTEQ4M parametrization \cite{CTEQ4} of parton 
densities, and present physical cross sections as a function of the 
$\tilde g$ mass or of the average mass 
$m = (m_{\tilde \chi} + m_{\tilde g})/2$.  For $p\bar p$ collisions at 
$\sqrt{S}=2 $ TeV the cross sections at $m_{\tilde g} = 300 $ GeV range from 
${\cal O}( 1{\rm pb} )$ for the $\tilde {\chi}^0_2$ and the 
$\tilde {\chi}^{\rm \pm}_1$ to ${\cal O}( 10^{-3}{\rm pb} )$ for the 
$\tilde {\chi}^0_3$.  The $\tilde g \tilde{\chi}^0_{(1,2)}$ and 
$\tilde g \tilde{\chi}^{\rm \pm}_1$ cross sections are of hadronic size 
despite the fact that the overall coupling strength is 
${\cal O}(\alpha_{EW}\alpha_s)$ not ${\cal O}(\alpha_s^2)$.  The 
masses of the $\tilde{\chi}^0_{(1,2)}$ and 
$\tilde g \tilde{\chi}^{\rm \pm}_1$ are significantly smaller in a typical 
mSUGRA scenario than those of the squarks and gluinos.  The phase space 
and the parton luminosity are therefore greater for associated production of a 
gluino and a gaugino than for a pair of squarks or gluinos, and the smaller 
coupling strength is compensated.  
The next-to-leading-order cross sections are enhanced by 
typically 10\% to 25\% relative to the leading order values.  The theoretical 
uncertainty resulting from variations of the factorization/renormalization 
scale is approximately $\pm 10\%$ at NLO for the 
$\tilde {\chi}^0_2$ and the $\tilde {\chi}^{\rm \pm}_1$, a factor of 2 
smaller than the LO variation.  Shown in Fig.~1 are the predicted cross 
sections as a function of the average mass.  

Baer, Harris, and Reno~\cite{Baer} compute total cross sections for all the 
slepton pair production reactions $\tell_L\tnu_L$, $\tell_L\bar{\tell_L}$,
$\tell_R\bar{\tell_R}$ and $\tnu_L\bar{\tnu}_L$ in next-to-leading
order QCD.  The analytic calculations are very similar to the QCD corrections
to the Standard Model massive lepton-pair production (Drell-Yan) process.
Numerical results are based on the CTEQ4M parametrization \cite{CTEQ4} of 
parton densities.   For $p\bar p$ collisions at $\sqrt{S}=2 $ TeV, the cross 
sections range from ${\cal O}( 1{\rm pb} )$ at $m_{\rm slepton}=50~$GeV 
to ${\cal O}( 10^{-3}{\rm pb} )$ at $m_{\rm slepton}=200~$GeV.  The 
next-to-leading-order cross sections are enhanced by typically 35\% to 40\% 
relative to the leading order values.  The theoretical uncertainty resulting
from variations in the hard scattering scale and parton distribution
functions is approximately $\pm 15\%$.
In the mSUGRA model, slepton pair production is most important for small
values of the parameter $m_0$.
The next-to-leading order enhancements of slepton pair cross sections 
at Tevatron energies can push predictions for leptonic SUSY signals to 
higher values than typically
quoted in the literature in these regions of model parameter space.

For current expectations of the hierarchy of masses and cross sections, 
consult Ref.~\cite{torah}.  

\section{Monte Carlo Methods}
\label{sec:2}

Experimental searches for supersymmetry rely heavily on Monte Carlo 
simulations of cross sections and event topologies. Two Monte Carlo generators 
in common use for hadron-hadron collisions include SUSY processes; they 
are ISAJET \cite{Paige}
and SPYTHIA \cite{Sjostrand,Mrenna}.  Both the Monte Carlo approach and the 
fixed order pQCD approach have different
advantages and limitations.  Next-to-leading order perturbative calculations 
depend on very few parameters, e.g., the renormalization and factorization 
scales, and the dependence of the production cross sections on these 
parameters is reduced significantly in NLO with respect to LO. 
Therefore, the normalization of the cross section can be calculated quite 
reliably if one includes the NLO contributions.  On the other hand, the 
existing next-to-leading order calculations provide predictions only for 
fully inclusive quantities, e.g., a differential cross section for 
production of a squark or a gluino, after integration over all other particles 
and variables in the final state.  In addition, they do not include sparticle 
decays. This approach does not allow for event shape studies nor for 
experimental selections on missing energy or other variables associated with 
the produced sparticles or their decay products that are crucial if one wants 
to enhance the SUSY signal in the face of substantial backgrounds from 
Standard Model processes.

The natural strength of Monte Carlo simulations consists in the fact that
they generate event configurations that resemble those observed in experimental 
detectors.  Through their parton showers, these generators include, in the 
collinear approximation, contributions from all orders of perturbation theory.  
In addition, they incorporate phenomenological hadronization models, a 
simulation of particle decays, the possibility to implement experimental cuts, 
and event analysis tools.  However, the hard-scattering matrix elements in 
these generators are accurate only to leading order in QCD, and, owing to 
the rather complex nature of infra-red singularity cancellation in higher 
orders of perturbation theory, it remains a difficult challenge to incorporate 
the full structure of NLO contributions successfully in Monte Carlo simulations.
The limitation to leading-order hard-scattering matrix elements leads to 
large uncertainties in the normalization of the cross section.  The parton 
shower and hadronization models rely on tunable parameters, another source of 
uncertainties.

In Ref. \cite{Berger} a method is suggested to improve the accuracy of the 
normalization of cross sections computed through Monte Carlo simulations.
In this approach, the renormalization and factorization (hard) scale in the 
Monte Carlo LO calculation is chosen in such a way that the normalization 
of the Monte Carlo LO calculation agrees with that of the NLO 
perturbative calculation.  The scale choice depends on which partonic 
subprocess one is considering and on the kinematics.  This choice of  
the hard scale will affect both the hard matrix element {\em and} the 
initial-state and final-state parton shower radiation.   On the other hand, 
an alternative rescaling of the cross section by an overall $K$-factor will 
have no bearing on the parton shower radiation.  A reduction in the hard scale 
leads generally to less evolution and less QCD radiation, and vice-versa, in 
the initial- and final-state showering.  A change of the hard scale will be 
reflected in the normalization of the cross section as well as in the event 
shape.  Investigations are underway to determine how significant the changes 
are in computed final state momentum distributions.  



\begin{figure}
 \begin{center}
  \vspace*{-10mm}
  \epsfig{file=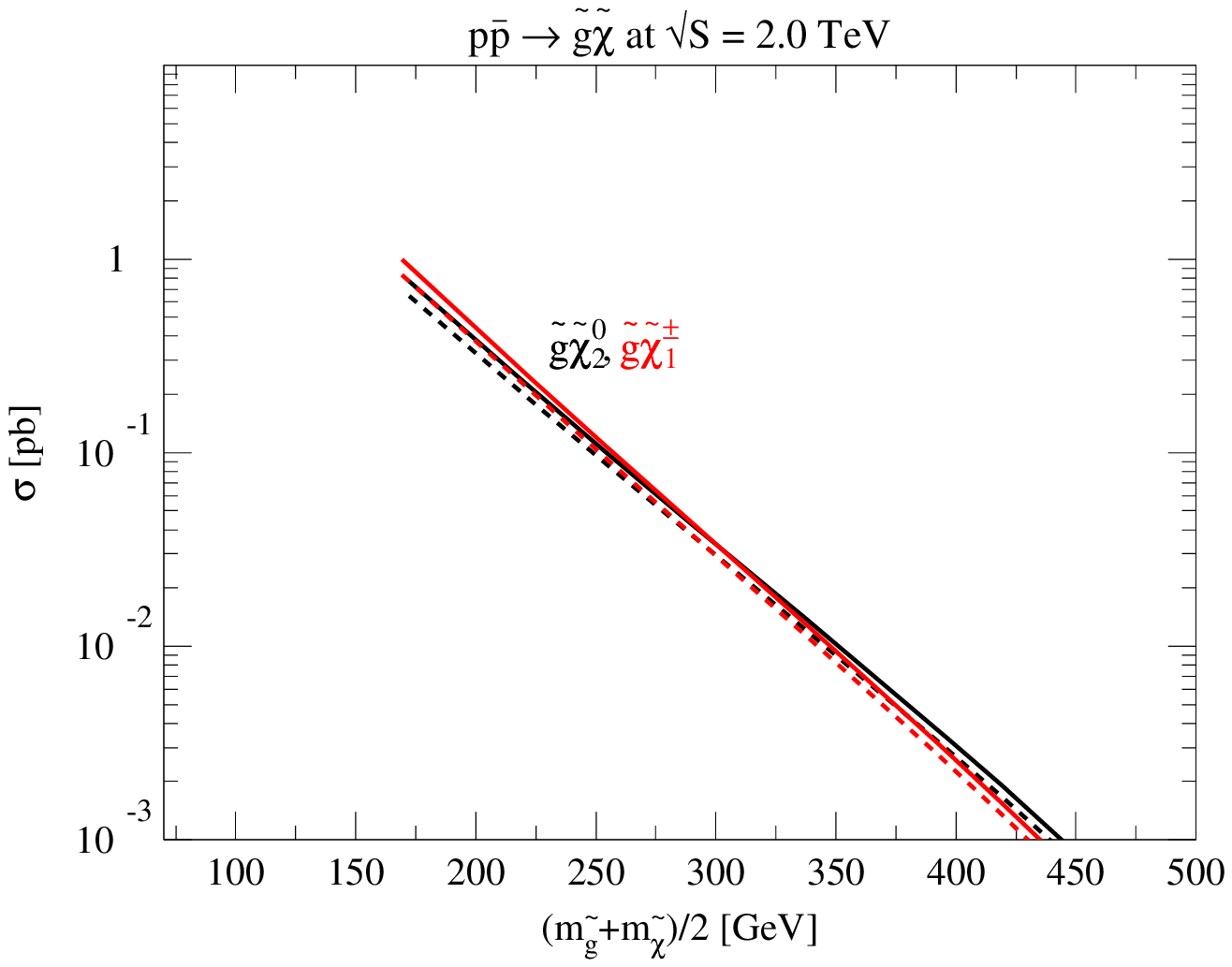,width=16cm}
  \vspace*{-4mm}
 \end{center}
 \caption{Total hadronic cross sections for the associated production
  of gluinos and gauginos at Run II of the Tevatron from Ref.[6]. 
  NLO results are shown as solid curves, and LO results as dashed curves.  
  The chargino cross sections are summed over positive and negative chargino 
  states.}
\label{xsec}
\end{figure}


\end{document}